\documentclass{article}
\usepackage{cite}
\usepackage{graphicx}
\usepackage{dcolumn}

\begin{document}

\title{Highly accurate calculation of the real and complex eigenvalues of
one-dimensional anharmonic oscillators}
\author{Francisco M. Fern\'{a}ndez \thanks{%
e--mail: fernande@quimica.unlp.edu.ar}\, and Javier Garcia \\
%EndAName
INIFTA (UNLP, CCT La Plata-CONICET), Divisi\'on Qu\'imica Te\'orica, \\
Blvd. 113 S/N, Sucursal 4, Casilla de Correo 16, 1900 La Plata, Argentina}
\maketitle

\begin{abstract}
We draw attention on the fact that the Riccati-Pad\'{e} method developed
some time ago enables the accurate calculation of bound-state eigenvalues as
well as of resonances embedded either in the continuum or in the discrete
spectrum. We apply the approach to several one-dimensional models that
exhibit different kind of spectra. In particular we test a WKB formula for
the imaginary part of the resonance in the discrete spectrum of a three-well
potential.
\end{abstract}

\section{Introduction}

\label{sec:intro}

In a recent paper Gaudreau et al\cite{GSS15} proposed a method for the
calculation of the eigenvalues of the Schr\"{o}dinger equation for
one-dimensional anharmonic oscillators. In their analysis of some of the
many approaches proposed earlier with that purpose they resorted to
expressions of the form: ``However, the existing numerical methods are
mostly case specific and lack uniformity when faced with a general
problem.'' ``As can be seen by the numerous approaches which have been
developed to solve this problem, there is a beautiful diversity yet lack of
uniformity in its resolution. While several of these methods yield excellent
results for specific cases, it would be favorable to have one general method
that could handle any anharmonic potential while being capable of computing
efficiently approximations of eigenvalues to a high pre-determined
accuracy.'' ``Various methods have been used to calculate the energy
eigenvalues of quantum anharmonic oscillators given a specific set of
parameters. While several of these methods yield excellent results for
specific cases, there is a beautiful diversity yet lack of uniformity in the
resolution of this problem.'' The authors put forward an approach that they
termed double exponential Sinc collocation method (DESCM) and reported
results of remarkable accuracy for a wide variety of problems. In fact they
stated that ``In the present work, we use this method to compute energy
eigenvalues of anharmonic oscillators to unprecedented accuracy'' which may
perhaps be true for some of the models chosen but not for other similar
examples. For example, in an unpublished article Trott\cite{T00} obtained
the ground-state energy of the anharmonic oscillator with potential $%
V(x)=x^{4}$ with more than 1000 accurate digits. His approach is based on
the straightforward expansion of the wavefunction in a Taylor series about
the origin.

One of the methods mentioned by Gaudreau et al\cite{GSS15} is the
Riccati-Pad\'{e} method (RPM) based on a rational approximation to the
logarithmic derivative of the wavefunction that satisfies a well known
Riccati equation\cite{FMT89a,FMT89b}. In their brief analysis of the RPM the
authors did not mention that this approach not only yields the bound-state
eigenvalues but also the resonances embedded in the continuum\cite{F95}.
What is more, the same RPM quantization condition, given by a Hankel
determinant, produces the bound-state eigenvalues, the resonances embedded
in the continuum as well as some kind of strange resonances located in the
discrete spectrum of some multiple-well oscillators\cite{F08}. It is not
clear from the content of\cite{GSS15} whether the DESCM is also suitable for
the calculation of such complex eigenvalues.

The accuracy of the calculated eigenvalues not only depends on the chosen
method but also on the available computation facilities and on the art of
programming. For this reason the comparison of the accuracy of the results
reported in a number of papers spread in time should be carried out with
care.

The purpose of this paper is two-fold. First, we show that the RPM can in
fact yield extremely accurate eigenvalues because it exhibits exponential
convergence. To that end it is only necessary to program the quantization
condition in an efficient way in a convenient platform. Second, we stress
the fact that the RPM yields both real and complex eigenvalues with similar
accuracy through the same quantization condition. More precisely: it is not
necessary to modify the algorithm in order to obtain such apparently
dissimilar types of eigenvalues that are associated to different boundary
conditions of the eigensolution.

In section~\ref{sec:RPM} we outline the RPM for even-parity potentials. In
section~\ref{sec:examples} we apply this approach to some of the examples
discussed by Gaudreau et al\cite{GSS15} and obtain eigenvalues with
remarkable accuracy. In this section we also calculate several resonances
supported by anharmonic oscillators that were not taken into account by
those authors. We consider examples of resonances embedded in the continuous
as well as in the discrete spectrum. Finally, in section~\ref
{sec:conclusions} we summarize the main results and draw conclusions.

\section{The Riccati-Pad\'{e} method}

\label{sec:RPM}

The dimensionless Schr\"{o}dinger equation for a one-dimensional model reads
\begin{equation}
\psi ^{\prime \prime }(x)+\left[ E-V(x)\right] \psi (x)=0,
\label{eq:Schrodinger}
\end{equation}
where $E$ is the eigenvalue and $\psi (x)$ is the eigenfunction that
satisfies some given boundary conditions. For example, $\lim\limits_{|x|%
\rightarrow \infty }\psi (x)=0$ determines the discrete spectrum and the
resonances are associated to outgoing waves in each channel (for example, $%
\psi (x)\sim Ae^{ikx}$). In this paper we restrict ourselves to anharmonic
oscillators with even-parity potentials $V(-x)=V(x)$ to facilitate the
comparison with the results reported by Gaudreau et al\cite{GSS15} but it
should be taken into account that the approach applies also to no
non-symmetric potentials\cite{F97}.

In order to apply the RPM we define the regularized logarithmic derivative
of the eigenfunction
\begin{equation}
f(x)=\frac{s}{x}-\frac{\psi ^{\prime }(x)}{\psi (x)},  \label{eq:f(x)}
\end{equation}
that satisfies the Riccati equation
\begin{equation}
f^{\prime }(x)+\frac{2sf(x)}{x}-f(x)^{2}+V(x)-E=0,  \label{eq:HO_Riccati}
\end{equation}
where $s=0$ or $s=1$ for even or odd states, respectively. If $V(x)$ is a
polynomial function of $x$ or it can be expanded in a Taylor series about $%
x=0$ then one can also expand $f(x)$ in a Taylor series about the origin
\begin{equation}
f(x)=x\sum_{j=0}^{\infty }f_{j}(E)x^{2j}.  \label{eq:f(x)_series}
\end{equation}
On arguing as in earlier papers we conclude that we can obtain approximate
eigenvalues to the Schr\"{o}dinger equation from the roots of the Hankel
determinant
\begin{equation}
H_{D}^{d}(E)=\left|
\begin{array}{cccc}
f_{d+1} & f_{d+2} & \cdots & f_{d+D} \\
f_{d+2} & f_{d+3} & \cdots & f_{d+D+1} \\
\vdots & \vdots & \ddots & \vdots \\
f_{d+D} & f_{d+D+1} & \cdots & f_{d+D-1}
\end{array}
\right| =0,  \label{eq:Hankel}
\end{equation}
where $D=2,3,\ldots $ is the dimension of the determinant and $d$ is the
difference between the degrees of the polynomials in the numerator and
denominator of the rational approximation to $f(x)$\cite
{FMT89a,FMT89b,F95,F08}. In those earlier papers we have shown that there
are sequences of roots $E^{[D,d]}$, $D=2,3,\ldots $ of the determinant $%
H_{D}^{d}(E)$ that converge towards the bound states and resonances of the
quantum-mechanical problem. We have at our disposal a set of sequences for
each value of $d$ but it is commonly sufficient to choose $d=0$. For this
reason, in this paper we restrict ourselves to the sequences of roots $%
E^{[D]}=E^{[D,0]}$ (unless stated otherwise).

In this paper we are concerned with anharmonic-oscillator potentials of the
form
\begin{equation}
V(x)=\sum_{j=1}^{K}v_{j}x^{2j}.  \label{eq:V(x)_AO}
\end{equation}
The spectrum is discrete when $v_{K}>0$ and continuous when $v_{K}<0$. In
the latter case there may be resonances embedded in the continuous spectrum
which are complex eigenvalues. The real part of any such eigenvalue is the
resonance position and the imaginary part is half its width $\Gamma $ ($%
\left| \Im E\right| =\Gamma /2$).

\section{Examples}

\label{sec:examples}

Four examples chosen by Gaudreau et al\cite{GSS15} are quasi-exactly
solvable problems; that is to say, one can obtain exact solutions for some
states:
\begin{equation}
\begin{array}{ll}
V_{1}(x)=x^{2}-4x^{4}+x^{6} & E_{0}=-2 \\
V_{2}(x)=4x^{2}-6x^{4}+x^{6} & E_{1}=-9 \\
V_{3}(x)=\frac{105}{64}x^{2}-\frac{43}{8}x^{4}+x^{6}-x^{8}+x^{10} & E_{0}=%
\frac{3}{8} \\
V_{4}(x)=\frac{169}{64}x^{2}-\frac{59}{8}x^{4}+x^{6}-x^{8}+x^{10} & E_{1}=%
\frac{9}{8}.
\end{array}
\label{eq:V_solvable}
\end{equation}
The RPM yields the exact result for all these particular cases because in
all of them the logarithmic derivative $f(x)$ is a rational function of the
coordinate. The Hankel determinants of lowest dimension for each case are:
\begin{eqnarray}
H_{2}^{0}(E) &=&{\frac{1}{4725}}\,\left( E+2\right) \left( {E}^{5}-2{E}%
^{4}-23{E}^{3}-602{E}^{2}+1030E-1412\right) ,  \nonumber \\
H_{2}^{0}(E) &=&{\frac{1}{4465125}}\,\left( E+9\right) \left( {E}^{5}-9\,{E}%
^{4}-187\,{E}^{3}-8217\,{E}^{2}+78336\,E-348624\right) ,  \nonumber \\
H_{3}^{0}(E) &=&{\frac{1}{3189612751764848640000}}\,\left( 8\,E-3\right)
\left( 8589934592\,{E}^{11}+3221225472\,{E}^{10}\right.  \nonumber \\
&&-1887235473408\,{E}^{9}-399347250364416\,{E}^{8}-1634745666502656\,{E}^{7}
\nonumber \\
&&+10770225531715584\,{E}^{6}-836065166572191744\,{E}^{5}  \nonumber \\
&&-905684630058491904\,{E}^{4}+5197219286067104256\,{E}%
^{3}-2944302537136698432\,{E}^{2}  \nonumber \\
&&\left. -12283878786837315912\,E+22452709866105906693\right) ,  \nonumber \\
H_{3}^{0} &=&{\frac{1}{431028319209742820966400000}}\,\left( 8\,E-9\right)
\left( 8589934592\,{E}^{11}+9663676416\,{E}^{10}\right.  \nonumber \\
&&-5569096187904\,{E}^{9}-2064531673055232\,{E}^{8}-15362232560910336\,{E}%
^{7}  \nonumber \\
&&+158709729905344512\,{E}^{6}-23752960275863896064\,{E}^{5}  \nonumber \\
&&-84068173973645402112\,{E}^{4}+2318080070178601634304\,{E}^{3}  \nonumber
\\
&&-6274577633554290840768\,{E}^{2}  \nonumber \\
&&\left. -75410626140297229262472\,E+655367638076442656931879\right) ,
\end{eqnarray}
respectively. It is clear that the second factor of each Hankel determinant
yields the exact eigenvalue of the corresponding model in equation (\ref
{eq:V_solvable}).

As a nontrivial example we consider the quartic anharmonic oscillator
\begin{equation}
V(x)=x^{2}+\lambda x^{4}.  \label{eq:V(x)_QAO}
\end{equation}
Gaudreau et al\cite{GSS15} calculated the ground state for $\lambda =1$ with
remarkable accuracy. The RPM also enables great accuracy because of its
exponential convergence. For example, with determinants of dimension $D\leq
623$ we obtained
\begin{eqnarray*}
E_{0} &=&1.3923516415302918556575078766099341846000667112208340889063493 \\
&&238775674318756465285909735634677917591211513753417388174455516 \\
&&240463837130438178697370013460935168154842085748896569018003055 \\
&&412366487432189534357154174093826240572295199985687111814096892 \\
&&270227363816981111260310703429386134195964568485918291463489851 \\
&&885814863025469392145221031177208948219643654580541741801366088 \\
&&701870825264349698158700823407607595743192268511389600196854493 \\
&&949820962407561620946196334634473774557014921149262346890591637 \\
&&338563062681405570992510627058090950578666603093583144835197352 \\
&&905560061049224302849821825415119194035000689109989896675454979 \\
&&833183805654199754661625730310527294045815675292625382286721180 \\
&&760183199752945956111132457567844565301841956779850974931537225 \\
&&418858821696022599972698095084658065637021365447651793869049904 \\
&&755455309191949465274340562585980971938979595684138772300267900 \\
&&681776732778457086544772456313662681845199346441260519691501249 \\
&&723061727243936387451149975151714249881364996642295004595485151 \\
&&916507248813368615814421881730600039773536840117104637678735672 \\
&&726392478420532548924901523470626991951440934018875830719295468 \\
&&178231131253774713120042218812766794224608722685106067661795491 \\
&&30792640798558850522732484547554994100518213983
\end{eqnarray*}
which is considerably more accurate than the result reported by those
authors. Such an accuracy is unnecessary but it clearly shows the stability
and remarkable rate of convergence of the RPM.

For some approaches the pure quartic oscillator
\begin{equation}
V(x)=x^{4}  \label{eq:V(x)=x^4}
\end{equation}
may be more demanding than the oscillator treated previously. However, this
is not the case for the RPM that yields
\begin{eqnarray*}
E_{0} &=&1.060362090484182899647046016692663545515208728528977933216245 \\
&&24169594356304434442112689629913467170351054624435858252558087 \\
&&98082102931470131768363738249357892262460047081754469601416374 \\
&&88417282256905935757790888061788790263601549395690275196148900 \\
&&94293487358440944269489790121397146429095192335453382834703350 \\
&&57576151120257039888523720240221841103086573731091398915453658 \\
&&41031116794058335486000922744006963112670238862297142969961059 \\
&&215583226671376935508673610000831830027517926233573913906136180 \\
&&776498596961814994127928092728407079561060440722946809949136275 \\
&&729273872791368902798424722261716944488954751370438068405439187 \\
&&787729532342458743725431783231906038106874160440343745301468472 \\
&&781391861294047043103401351071607110353008929823275427661518986 \\
&&950565047160252756089526262191025688200964410287815640052705292 \\
&&932405076382650282591124773625384718547144025722854384852974504 \\
&&585709788402490669995704768445877091762029124375273254907116433 \\
&&440230294730692398190895685374535988446016002313291933059395869 \\
&&304916644281633946163324287004261461237743009952234204208597735 \\
&&690153565416850308941851348795734106585479719467596466796613467 \\
&&688586437952654519560568286715958338884743467012042420714918747 \\
&&871038429573389138985245894022263471696176996560440931170998547 \\
&&160646641857421281143088181114951122148431408871216620593130769 \\
&&234180229827246883626045356507913236221596486925870033200744409 \\
&&688064046239788178394698378070482686021742719460350750696191658 \\
&&224983009606134572666392863592176435340137189204481484648373028 \\
&&941252963863440446954353934473733433447707230478215508820964235 \\
&&1106900382833900237848230939194834
\end{eqnarray*}
from determinants of dimension $D\leq 806$. We think that present result is
more accurate than the one reported by Trott\cite{T00}, the discrepancy
being in his last 9 figures. In the case of the quartic anharmonic
oscillators (\ref{eq:V(x)_AO}) and (\ref{eq:V(x)_QAO}) it was proved that $%
E_{0}^{[D,0]}$ and $E_{0}^{[D,1]}$ are lower and upper bounds to the actual
eigenvalue, respectively\cite{FMT89a,FMT89b}. In order to verify the
accuracy of present calculation we verified that both bounds agreed to the
last digit.

As stated in the introduction, the RPM yields not only the bound states but
also the resonances. For example, in the case of the anharmonic oscillator (%
\ref{eq:V(x)_QAO}) with $\lambda =-0.1$ (inverted double well) we obtain
\begin{eqnarray*}
\Re E &=&0.900672904092015024804721689210287758304603316620306983171851 \\
&&6924034871463944702496167267900896286882529377637746473257725 \\
&&7777502973177491351445744733858777092680397126780146975586991 \\
&&7579845522510002054201377209951767663713270344807430710771398 \\
&&0087908739856469918617515352140215990178673201060287973953945 \\
&&7665858699755629731892792582490816179032126417574633154406875 \\
&&6411634377154244333590011870423051096533092590779249514766962 \\
&&8533284309302112344770277963917083285621130844172690957306044 \\
&&2388660682777957152776169280604258100975307902890897826798321 \\
&&3678426826574845962017573001105365337069958536171568094454228 \\
&&5361299988968687820123088536512736472768963293804541994622283 \\
&&1027030494463591543444008242687831194918572931500609956657810 \\
&&8882411457777224871083716564437160787964973794652063487913034 \\
&&212377042606399056515807797185750609934729619
\end{eqnarray*}
and
\begin{eqnarray*}
\left|\Im E\right|
&=&0.006693280875800130269271875081318241122949894326169673589331 \\
&&4097282610915605850430183935416396746724362148135740974410006 \\
&&9700186351017147154410094496471209525956636119425938632579451 \\
&&9366933621549986695707277857784464014031443123369559867398037 \\
&&5835418405468882108847886624888017187187133797463636836846120 \\
&&5368517345681897772416233328036067770330148991244629883789645 \\
&&5581815166460445660555875185436390373393566787035417128350480 \\
&&8639418360504495322170748966013843419825115259187606452383097 \\
&&8320147077328407782186985142956414437484959838166597216101069 \\
&&5961186324314282177836539138159742843180320299235738874296270 \\
&&5328614872135008968453119432810683411008857433705994892657399 \\
&&1903248493789405585502725824263568102961538618496687088553748 \\
&&0837946534215566858502262251410323283978252816367927891368870 \\
&&401758500263860266656837355486939353652027140
\end{eqnarray*}
with determinants of dimension $D\leq 429$. As far as we know this resonance
was not calculated with such an accuracy before.

\begin{figure}[tbp]
\begin{center}
\includegraphics[width=9cm]{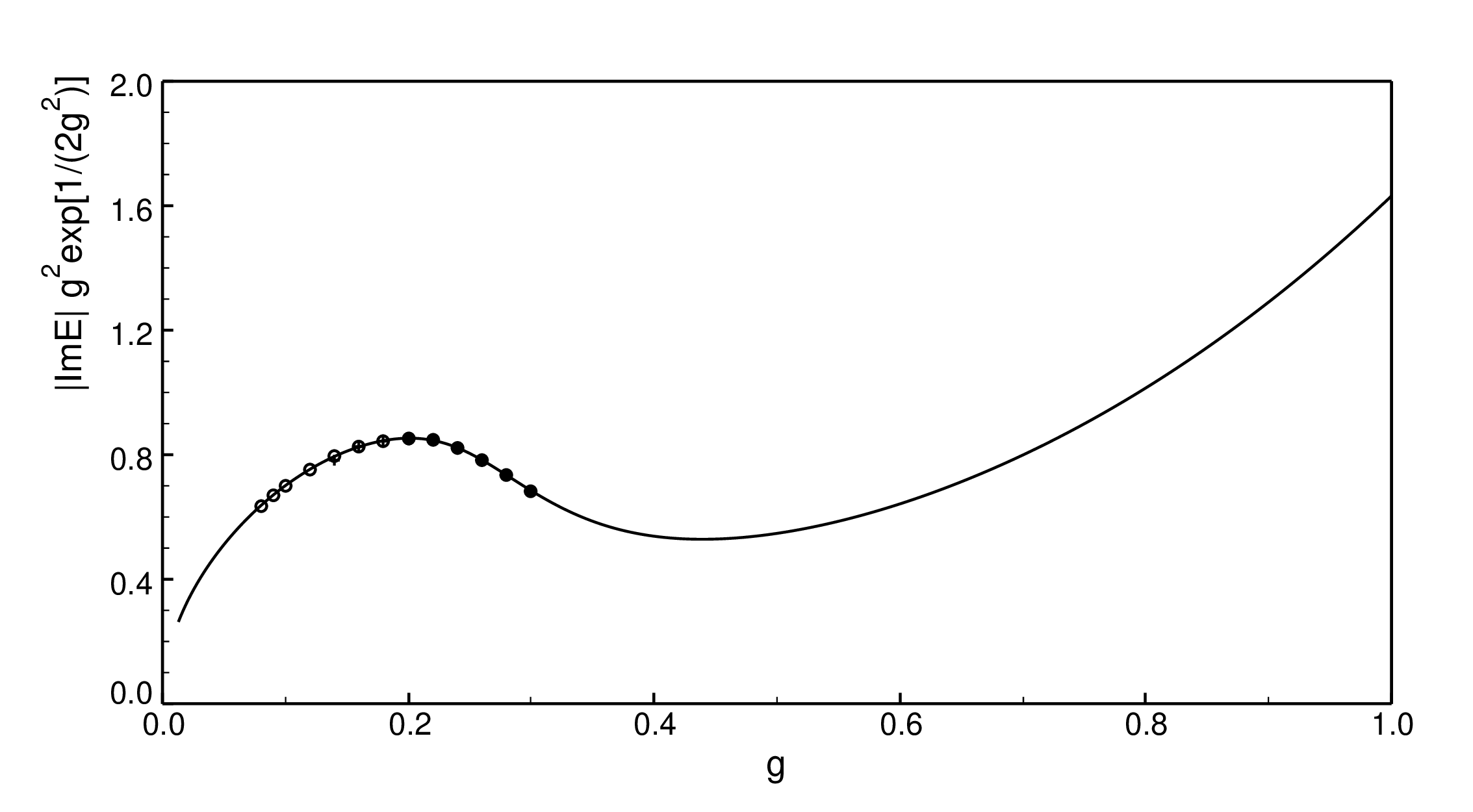}
\end{center}
\caption{Present calculation of $\left|\Im E\right|\ g^2 \exp\left[1/\left(2
g^2\right)\right]$ for the oscillator (\ref{eq:V(x)_TW}) (solid line) and
the results of Benassi et al\protect\cite{BGG79} (filled circles),
Killingbeck\protect\cite{K07} (crosses) and Fern\'andez\protect\cite{F08}
(empty circles).}
\label{fig:TWWKB}
\end{figure}

A most interesting example is given by
\begin{equation}
V(x)=x^{2}\left( 1-g^{2k}x^{2k}\right) ^{2},\;k=1,2,\ldots .
\label{eq:V(x)_TW}
\end{equation}
It is a three-well potential with minima $V(x_{m})=0$ at $x_{m}=-1/g,0,1/g$.
Since $V(x)$ behaves asymptotically as $g^{4k}x^{4k+2}$ when $|x|\rightarrow
\infty $ then one expects only bound states with positive eigenvalues when $%
g $ is real. However, Benassi et al\cite{BGG79} proved that this family of
potentials supports complex eigenvalues with all the properties of actual
resonances. They calculated the lowest resonance for $k=1$ and several
values of $g$ and compared the imaginary part with the WKB transmission
coefficient through the barrier $\left| \Im E_{WKB}\right| \sim
Ag^{-2}e^{-1/(2g^{2})}$. Later Killingbeck\cite{K07} and Fern\'{a}ndez\cite
{F08} calculated this resonance more accurately and for smaller values of $g$
and showed that $\left| \Im E\right| \ g^{2}e^{1/(2g^{2})}$ does not exhibit
a uniform behaviour as suggested by the earlier calculation of Benassi et al%
\cite{BGG79}. In particular, Killingbeck\cite{K07} suggested that $\left|
\Im E\right| \ g^{2}e^{1/(2g^{2})}$ exhibits a maximum. In this paper we
calculated $\Im E$ even more accurately and for smaller values of $g$ and
present results, shown in Fig.~\ref{fig:TWWKB}, suggest that the conjecture
that ``the imaginary part of the resonance behaves as the WKB transmission
coefficient through the barrier'' may not be correct. The results of Benassi
et al\cite{BGG79} for a shorter interval of $g$ (also shown in the figure)
give the impression that $\left| \Im E\right| $ has already reached the
asymptotic behaviour $Ag^{-2}e^{-1/(2g^{2})}$ which is not the case. The
figure also shows the earlier results of Killingbeck\cite{K07} and
Fern\'{a}ndez\cite{F08}. A straightforward least-squares fitting of present
results suggests that the correct behaviour is $\left| \Im E\right| \sim
Ag^{-3/2}e^{-1/(2g^{2})}$.

\begin{figure}[tbp]
\begin{center}
\includegraphics[width=9cm]{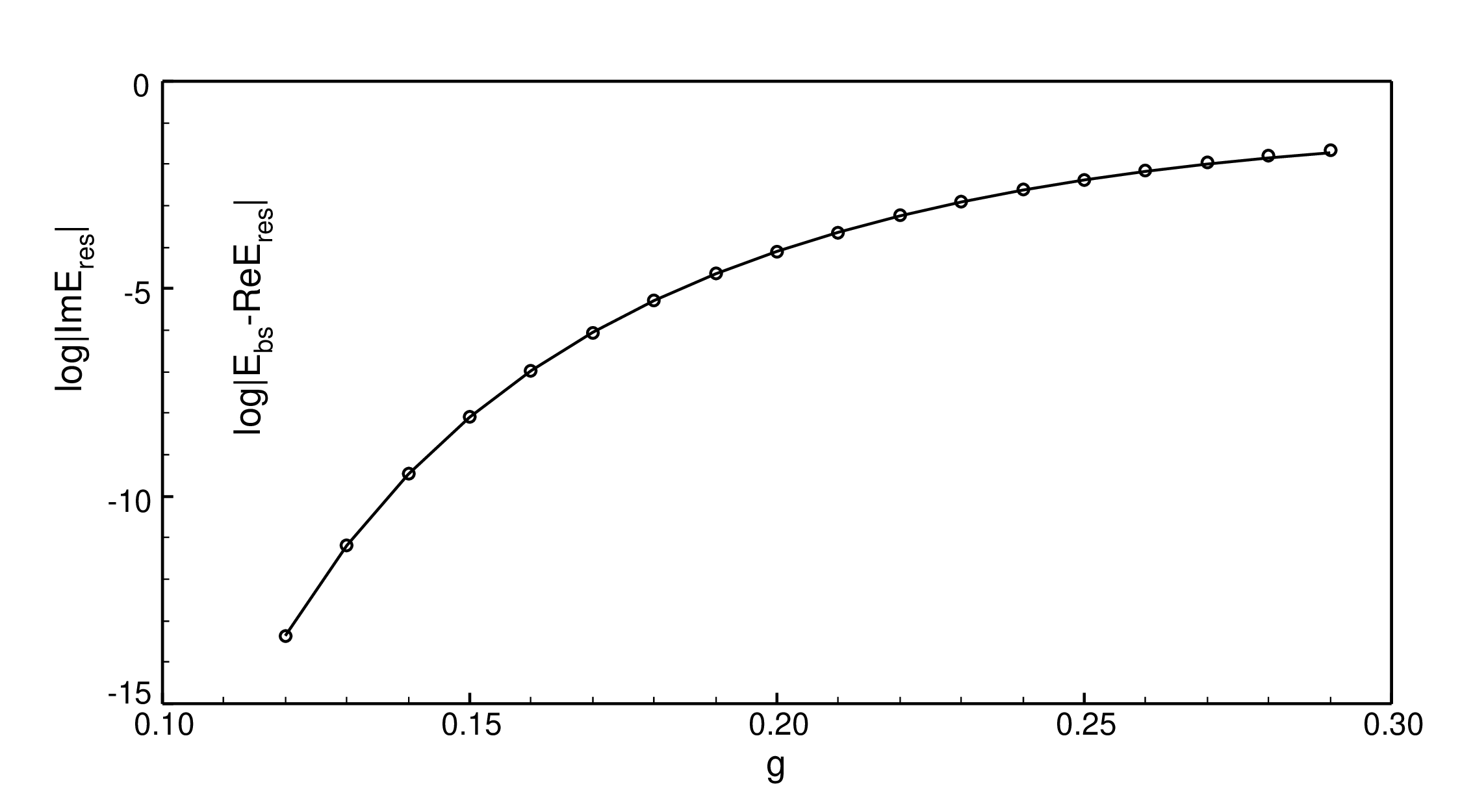}
\end{center}
\caption{$\log\left|\Im E_{res}\right|$ (circles) and $\log\left|\Re
E_{res}-E_{bs}\right|$ (solid line) vs. $g$ for the oscillator (\ref
{eq:V(x)_TW}) with $k=1$.}
\label{fig:TWREIM}
\end{figure}

The lowest bound state $E_{bs}$ and the real part of the lowest resonance $%
\Re E_{res}$ approach each other as $g\rightarrow 0$ in such a way that $%
\left| E_{bs}-\Re E_{res}\right| $ is of the order of $\left|\Im
E_{res}\right|$. This fact is clearly shown in Fig.~\ref{fig:TWREIM}.

The RPM enables us to calculate the bound states and resonances quite
accurately. In what follows we show some of them for the potential (\ref
{eq:V(x)_TW}) with $k=1$ and $g=0.2$.

By means of the RPM for even solutions and from determinants of dimension $D
\leq 775$ we have estimated

\begin{eqnarray*}
E_{bs} &=&0.932476291964221250713283307051702588320858910 \\
&&16450940452195530440541397037312615064902568796 \\
&&08959269025187554700854590142048116044798515032 \\
&&74607576315694596343111670051681020512461078632 \\
&&25947411410076408768061413506747150659467931140 \\
&&45773553564796578081916410057605227149530113848 \\
&&32488596806119963850647806580001262138356174439 \\
&&92627876704264450003234826765601226526080854117 \\
&&18883901908126394568677114253841158525832058185 \\
&&89486604967394261040369468954266459806798388982 \\
&&94552842394543958516859760723678410546834167750 \\
&&46265615981244952016961806157227974852224223062 \\
&&63307833915772881280212070865385935277969972309 \\
&&73792177447292258928727153726473008063005886824 \\
&&27244516489544916400797944748255960907744830950 \\
&&81889615946058064623691858375390646083072560381 \\
&&93215181635793814371703342844510993772509125579 \\
&&56334986610268220762863856194297627513825342515 \\
&&50679107510710529342767883873476401983945574715 \\
&&11757502147307390721534786949037416819236026830 \\
&&61521556243352973248410809257466846919735250978 \\
&&076767512840642337653806458904012731861910502
\end{eqnarray*}
Analogously for the lowest resonance and from determinants of dimension $%
D\leq 691$ we obtained:

\begin{eqnarray*}
\Re E_{res} &=&0.93255571582477452179676759062168990966452649573413 \\
&&85221674869167644974697878297482889666009195213742 \\
&&93717600069469906629572410914840067688495198625372 \\
&&17985690364409228164242506169354102397596504534361 \\
&&60103546722503111068374776347405701051766932138082 \\
&&85413600743099773501734379824559724078809846013142 \\
&&54466796141742916190093031117808393336341712037560 \\
&&74384668770561453730731352495106674855706489027647 \\
&&57628951678813837141930823450477006238516389060959 \\
&&38361501163923123468457211692027452357327627884344 \\
&&99112499793762153016569474982638402568720610449675 \\
&&76070472867118837391778386061630035039234021063388 \\
&&55589439873306969552123357849387411686488958371712 \\
&&88731653433415162889450633825069899826979320533720 \\
&&96508988556846872196309982663487618177697150533924 \\
&&72202304048467020861359787434424955986376342243023 \\
&&12693029687970276633152983867273666099801795397355 \\
&&82538737743202827605750205148961320017464691543602 \\
&&10909829129113885230560751170580075050547310634884 \\
&&89167352701182508545578769602119439587479454959070 \\
&&96406141902643986905961993843552992164570091818648 \\
&&46232
\end{eqnarray*}
and

\begin{eqnarray*}
\left|\Im E_{res}\right|
&=&0.0000794775543996767650576037789218578987811077421 \\
&&147849846562298140940287872947528146204563615614090 \\
&&028774445460294456021141163045800235709638364949663 \\
&&353905939799131542366777545481790495539496899144485 \\
&&754062579429110541064567909497805641816263780566675 \\
&&388233499495422056087561163648183005687447211370677 \\
&&159865123112232223161645886340846246513728783929916 \\
&&686176186086127723970824686813236167800836809517307 \\
&&934131065936050216138586041138997060636301660413252 \\
&&566742473343264345284070048369409457383472779415364 \\
&&601962826737820988546159296790667149898273960541718 \\
&&035505319649723627339124594304665614834862888093604 \\
&&238202929173352801815954037592272451961628205950969 \\
&&404764180454049311285199354566868017448385193114605 \\
&&883829806930057675979698384480925313088391011486261 \\
&&124472043838440003626344405429021299123105595053241 \\
&&661288059523168203490179201354209577049949175017998 \\
&&930282112536345053702570708768896012135190244497127 \\
&&005841328004652182923323394262461014032693652500467 \\
&&599419583388579441325365273718235875966252696289277 \\
&&9223128751255675814299886872626419368
\end{eqnarray*}
There are also resonances among the odd solutions that, as far as we know,
have not been reported so far. For example, from determinants of dimension $%
D\leq 675$ we have estimated

\begin{eqnarray*}
\Re E &=&2.61567434444732550869411505135245734080470814119 \\
&&8081336401725389930302231715543431391542639715932 \\
&&6112732616783541689552149339774924576459116223527 \\
&&4031645919333724417808271441248941817228663854017 \\
&&1598561805645280654160019475399671189366467557187 \\
&&7023231933795457998655216288018422680281312142150 \\
&&0900102213105491896812637673117118610977610908627 \\
&&3339313659663532602566940509690710619685623673577 \\
&&4702051685651830157091563631235768241343388226233 \\
&&4214966804682833763697032153747063887197465955131 \\
&&0578860140998664952511675889072946426592233312325 \\
&&6683696960592694142085005517688414623839393719684 \\
&&4660149637044379455905460056628569763686110064049 \\
&&1982394293180159214983414747793630741582080077825 \\
&&7407910054254871310091689731748516412820369579467 \\
&&3252836879151952555347108138232897278070434422830 \\
&&0626704068364463061804095353910257042206327201629 \\
&&1297397160490735553290365416147019439703881706172 \\
&&7015605838456405956320547199197731222371367961928 \\
&&2365097186125212629398582930268581704082229677155 \\
&&6014934576275263959253938748903578577589148738386 \\
&&671
\end{eqnarray*}
and

\begin{eqnarray*}
\left|\Im E\right| &=&0.012103006054949689419532092547676879902909061831561
\\
&&14606655397483574310015080238580046511006197881958452 \\
&&80231225742919968027487497671478032524753374025385051 \\
&&78880120535081995629925326152262013101829054124399674 \\
&&83243664748674923025125583004541265899351700981055825 \\
&&50256507625482608264912298496868223692732502872218110 \\
&&91176377769035782652477078300369066474079388696220057 \\
&&74215632091529416042979012944093555392893071799032327 \\
&&21645370003682922872620954763793543027651537371155188 \\
&&04371553808569321951667977345154403280617508551577861 \\
&&29445926375665988730330512036322261513744864521271324 \\
&&70480449317847540296878526243141260350713152579083265 \\
&&00685413450993979606171778329687752598902925556571014 \\
&&20400501502261841417315902328430591349050070609689513 \\
&&88654575917732480890069301725502843827758335258773307 \\
&&44338536629042846929583130852297856757568562796662165 \\
&&61614392371093267546905111156935850217393851260547165 \\
&&10517706462696620628729983577471054074553526126483662 \\
&&11910087383972814671824614174682165252936046723308704 \\
&&7280261989144813328509313
\end{eqnarray*}

\begin{figure}[tbp]
\begin{center}
\includegraphics[width=9cm]{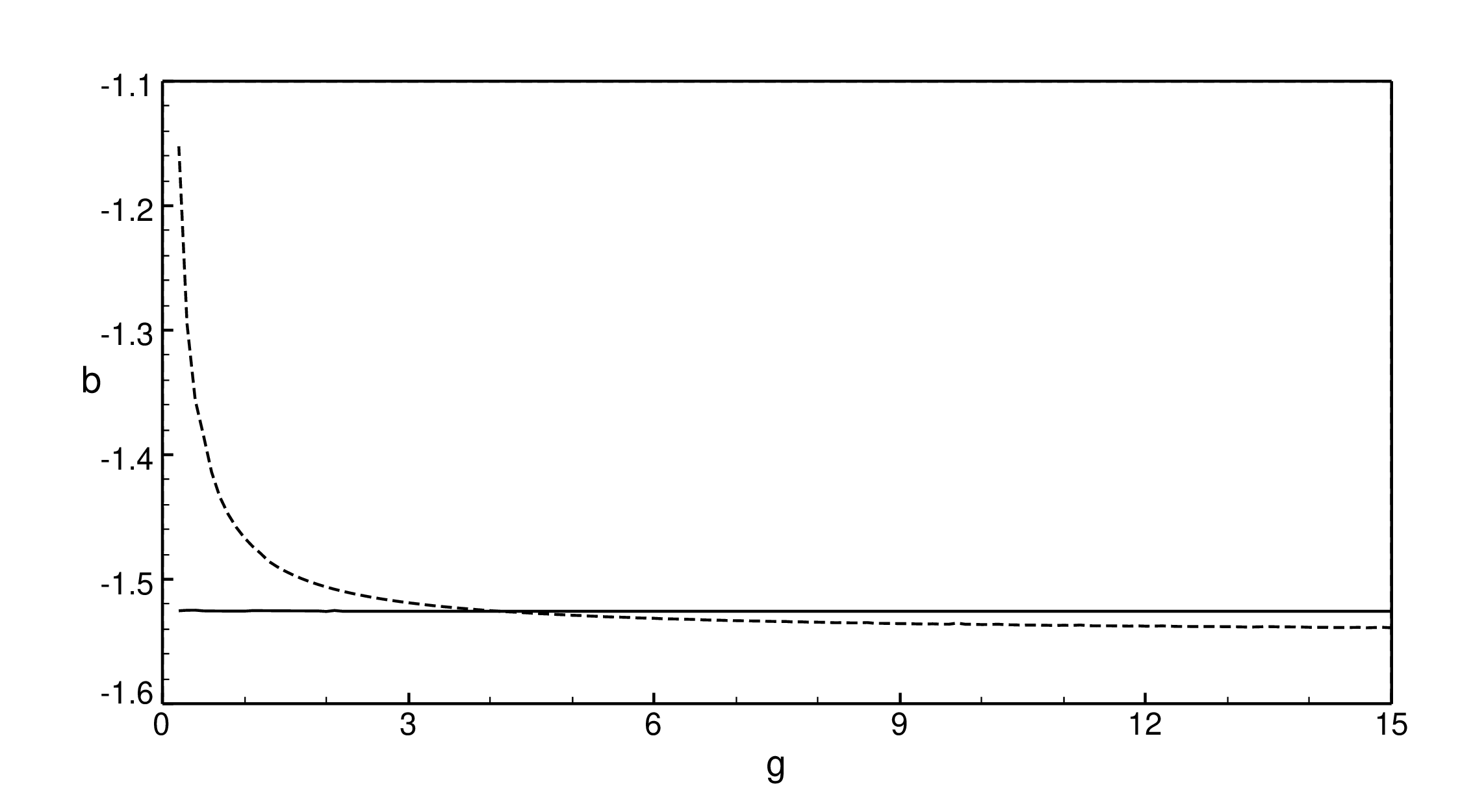}
\end{center}
\caption{Slope $b(g)$ for the lowest even bound state (dashed line) and
resonance (solid line) of the oscillator (\ref{eq:V(x)_TW}).}
\label{fig:TWSLOPE}
\end{figure}

The rate of convergence of the RPM for the bound states and the resonances
may be different. One way of monitoring the rate of convergence is to
calculate $\log \left| E^{[D]}-E^{[D-1]}\right| $ which is a straight line
when the rate of convergence is exponential. In the present case we fit $%
y(D)=a+bD$ to $\log \left| E^{[D]}-E^{[D-1]}\right| $ and obtain $b(g)$ for
several values of $g$. The results for the lowest even bound state and
resonance are shown in figure~\ref{fig:TWSLOPE}. That figure clearly shows
that the rate of convergence for the resonance is almost constant whereas it
decreases with $g$ for the bound state. As a result it is possible to obtain
the resonance for small values of $g$, say $g<1$, more accurately by means
of determinants of similar and even lesser dimension. However, this
advantage is counterbalanced by the fact that the mathematical operations
require more CPU time when complex numbers are involved.

In order to carry out the necessary arithmetic of complex numbers with
arbitrarily high precision we resorted to the GNU MPC C library\cite{AGTZ15}
and to the recurrence relation:
\begin{equation}
H_{D}^{d}=\frac{H_{D-1}^{d}H_{D-1}^{d+2}-\left( H_{D-1}^{d+1}\right) ^{2}}{%
H_{D-2}^{d+2}},  \label{eq:Hankel_rec_rel}
\end{equation}
for a fast calculation of the Hankel determinants.

\section{Conclusions}

\label{sec:conclusions}

Throughout this paper we tried to stress two points. The first one is that
the RPM can yield eigenvalues of remarkably accuracy if the algorithm is
programmed judiciously. To this end we have calculated the lowest
eigenvalues of the oscillators (\ref{eq:V(x)_QAO}) (with $\lambda =1$) and (%
\ref{eq:V(x)=x^4}) with greater accuracy than those reported by Gaudreau et
al\cite{GSS15} and Trott\cite{T00}, respectively.

The second point is that only one RPM quantization condition applies to
bound states and resonances. To illustrate it we calculated the resonances
for two models with great accuracy. One of them is an ordinary resonance
embedded in the continuum and other one is some kind of strange resonance
appearing in the point spectrum close to the ground state. Present results
for the lowest resonance in the discrete spectrum of the three-well
potential (\ref{eq:V(x)_TW}) give support to the conjecture that the
analytical WKB formula for the resonance width derived several years ago\cite
{BGG79} may not be correct. Present results improve considerably upon those
reported earlier by Killingbeck\cite{K07} and Fern\'{a}ndez\cite{T00}. It is
not clear to us whether the DESCM\cite{GSS15} or the power series approach%
\cite{T00} may also be applied to resonances without considerable
modification of the calculation algorithms.

It is not our purpose to criticize the DESCM which is clearly a
powerful algorithm as already proved by the remarkable
calculations carried out by Gaudreau et al\cite{GSS15} on a wide
variety of one-dimensional models. We just wanted to draw
attention to some advantages of the RPM that have been overlooked
in the discussion of the method carried out by those authors.

\end{document}